# Emerging Platform Work in the Context of the Regulatory Loophole (The Uber Fiasco in Hungary)

Csaba Makó[1] – Miklós Illéssy[2] – József Pap[3] Saeed Nostrabadi[4]

**Abstract**

The study examines the essential features of the so-called platform-based work, which is rapidly evolving into a major, potentially game-changing force in the labor market. From low-skilled, low-paid services (such as passenger transport) to highly skilled and high-paying project-based work (such as the development of artificial intelligence algorithms), a broad range of tasks can be carried out through a variety of digital platforms. Our paper discusses the platform-based content, working conditions, employment status, and advocacy problems. Terminological and methodological problems are dealt with in-depth in the course of the literature review, together with the 'gray areas' of work and employment regulation. To examine some of the complex dynamics of this fast-evolving arena, we focus on the unsuccessful market entry of the digital platform company Uber in Hungary 2016 and the relationship to institutional-regulatory platform-based work standards. Dilemmas relevant to the enforcement of labor law regarding platform-based work are also paid special attention to the study. Employing a digital workforce is a significant challenge not only for labor law regulation but also for stakeholder advocacy.

Keywords: Platform works, digital labor, work force, digital platform, digitalization

Journal of Economic Literature JEL codes: J21, M54, O33

[1] Csaba Makó is a professor emeritus, University of Public Service, 1083 Budapest, (e-mail: mako.csaba@tk.mta.hu).
[2] Miklós Illéssy is a researcher Center for Social Sciences– Hungarian Academy of Sciences, Center of Excellence, Budapest.
[3] József Pap is a PhD student, Széchenyi University Doctoral School of Management (SzEEDSM) Győr.
[4] Saeed Nostrabadi is a PhD Candidate, Hungarian University of Agriculture and Life Sciences, Gödöllő, (e-mail: saeed.nostrabadi@gmail.com).

### i. Introduction

To put it quite simply, today we are forced to cope with three mega crises: the Covid-19 pandemic and its socio-economic and cultural implications that are difficult to measure; the short- and long-term threats of climate change; and the disruptive effects of digital technology on the labor market (artificial intelligence, robotics) from the perspectives of the Fifth Industrial Revolution (Mazzucato, 2020). The social and economic players of each national economy are forced to tackle the historical challenges of social and economic regulation in the light of the Triple Crisis, the root of which is the change in the technical and economic paradigm (Perez, 2010).

The emergence of the digital economy, in which the secret to success is not ownership of physical resources but control of networks and intangible assets, is one of the most important new aspects of $21^{st}$-century capitalism. The change in the ranking of the world's most profitable businesses between 2008 and 2018 reflects this fundamental change. Whereas the energy and telecommunication giants dominated among the most profitable firms in the world in decades past, today the digital data industry's mammoths have taken their place.

**Table 1: Largest US companies in 2008 vs 2018**

| Rank | Company | Founded | US bn | Company | Founded | US bn |
|---|---|---|---|---|---|---|
| | | 2008 | | | 2018 | |
| 1. | Exxon | 1870 | 492 | Apple | 1976 | 891 |
| 2. | General Electric | 1892 | 358 | Google | 1998 | 768 |
| 3. | Microsoft | 1975 | 313 | Microsoft | 1975 | 680 |
| 4. | AT&T | 1885 | 238 | Amazon | 1994 | 592 |
| 5. | Procter & Gamble | 1837 | 226 | Facebook | 2004 | 545 |

*Source:* own compilation based on Johnston (2018)

As can be seen from Table 1, only Microsoft and AT&T were among the top 5 IT companies in 2008. Ten years later, all five of the largest companies emerged from

Electronic copy available at: https://ssrn.com/abstract=3764851

the digital economy, and while three are clearly platform companies[5], Apple and Microsoft still benefit significantly from and support the platform economy. These platforms differ in a variety of ways: there are primarily platforms that thrive by selling advertisement space using targeted personal data (such as Facebook, Google), there those that provide cloud services (such as Amazon Web Services), there are others that sell between a range of clients and users (such as Amazon, Airbnb), there are job placement platforms (such as Uber, Amazon Mechanical Turk, Upwork), and more recently, there are industrial platforms (such as Siemens, Jabil).

In particular, with the social transition powered by digitalization, a broad variety of markets are fundamentally altered by services (like Uber), goods (such as eBay), video-based content (such as YouTube), finance (such as Prosper), and the labor market (Upwork). Uber, for instance, "…converts taxi company employees or former medallion owners into contractors, whose access to income is through the Uber platform, while removing the government from the rate-setting equation." (Kenney and Zysman, 2016: 9)

The company is a clear example of the platform sector's extremely fast growth. Established in San Francisco in 2009, within five years after its establishment the organization has provided services in hundreds of cities in six countries, with an estimated value of $60 billion by 2016 (Kovaleski, 2020). While Uber and other companies using digital platforms currently account for a small fraction of the total workforce according to reliable estimates, this form of employment is rapidly increasing (Schwellnus et al., 2019: 8). It can be anticipated that the spread of the coronavirus outbreak would stimulate more job and work development in the digital labor markets.

---

[5] In the sense that the share of the goods they own is negligible compared to the share of the goods they move.

Analyzing this sector presents some substantial challenges. As the majority of platform companies operate in the 'gray zone' of institutional regulation, it is difficult to reliably identify the number of participants on the basis of the data available. In 2018, an annual income of $100,000 or more was registered by 3.3 million high-wage independent platform employees, and a survey conducted before the Covid-19 showed that the share of platform employees was on the rise (Duszynszki, 2020: 9-10).

A number of surveys in Europe have attempted to estimate the proportion of platform workers, the findings of which are difficult to compare due to methodological differences. Huws et al. (2019) for example, indicated prevalence rates of 5 to 20 percent in their research covering 13 European nations. The COLLEEM research (Pesole et al., 2018) found much lower rates: generally, the number of workers receiving monthly revenue from a platform work varies from 5 to 10 percent. These two surveys estimated the proportion of those who earn at least half of their revenue from the platform work at 0.5 to 6.5 percent. The *European Trade Union Institute* (ETUI) surveyed the prevalence of platform work in post-socialist countries: in Hungary, 7.8 percent of workers rarely work through platforms, 3 percent at least monthly, 1.9 percent work regularly on platforms, and 3.4 percent earn at least half of their earnings from platforms (Piasna and Drahokoupil, 2019: 16). But there is emerging evidence that the overwhelming majority of platform workers in the United States and elsewhere have become virtually unemployed and may suffer from depression since the emergence of the coronavirus and related economic impacts that have resulted in a large proportion of them not being eligible for unemployment benefits or other wage subsidies despite being employed in the so-called contractual form (Blusteinet al., 2020).

Our paper is one of the first products of an international research launched in 2020. The *Crowdwork21* project was supported by the *Directorate-General for Employment, Social Affairs and Inclusion* (DG EMPL) in the EU. The objective of the research is to examine the impact of platform-based work on employment, working conditions and social dialogue at work in particular. We have reviewed foreign and domestic literature in the first phase, and our own field research takes place at the time our study is published. However, because of the novelty of the topic, we thought it was worthwhile at this early stage to present a short article on this type of work because it poses important challenges not only for employees and trade unions, but also for traditional employers and all stakeholders involved in public policymaking.

In the digital network economy, the characteristics of work and employment cannot be explained solely by technological processes. In this context, Grabher and Tuijl (2020) emphasizes: "platform operators are not simply match-makers but instead veritable market-makers. As market-makers, platform operators not only enable individual transactions but actually frame the entire institutional and regulatory framework of the platform economy" (Grabher and Tuijl, 2020: 1012.).

We illustrate through the transport distribution companies Uber and Bolt that the growth of the institutional and regulatory system is not really dictated by technical development, but by the social actors involved.

In this paper, we first briefly examine the terminological discussions and the key terms used in connection with platform-based work. Through the example of Uber's failed attempt to enter Hungary and the resulting aftermath, which resulted in the alternative platform company now known as Taxify, we will then present the regulatory challenges of platform-based work. We summarize the key elements of the debate on the subject of algorithm-based management in the Hungarian labor law community and its problems in labor law. Finally, future theoretical and methodological research problems are highlighted.

**ii. Methodology**

The research method of the present study has been designed to classify and review notable articles in the development of the concept of platform work. The academic literature, along with gray literature, and employment regulations were used as data sources for this study. The Elsevier and Scopus databases were used to search for scientific articles. The research method uses a comprehensive and structured mechanism based on a systematic database search and gray literature. Figure 1 shows the process of conducting the research method.

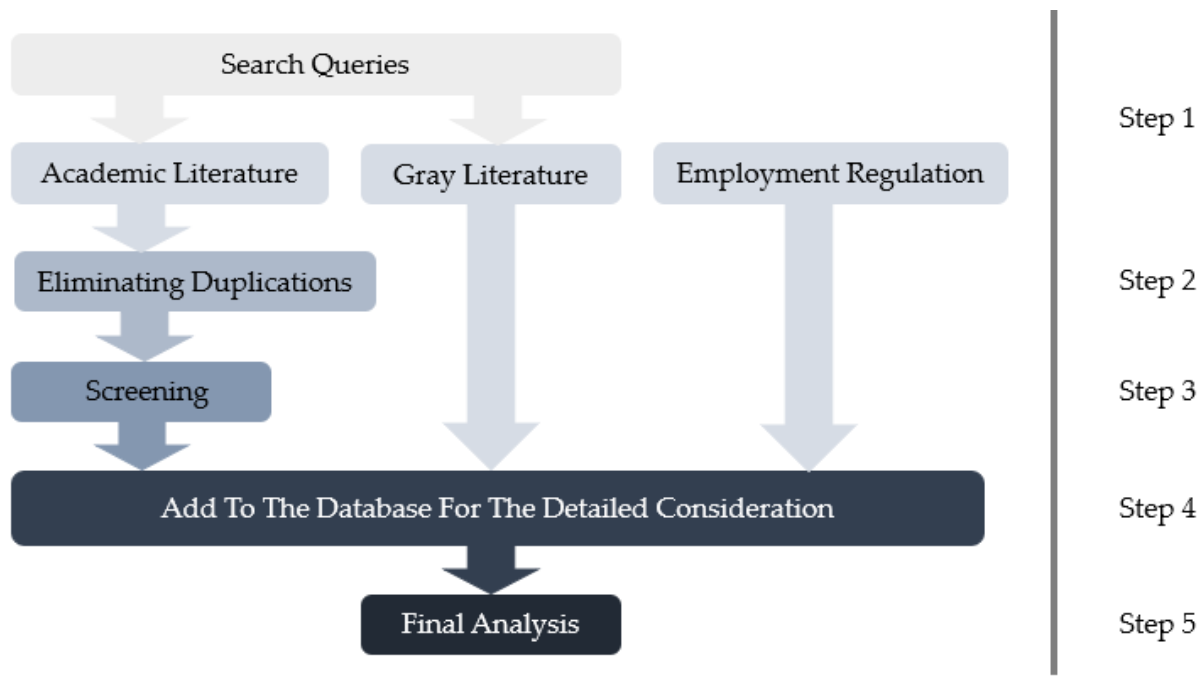

**Figure 1.** Diagram of the methodology of research.

To achieve the objectives of this study, structured and comprehensive processes were designed in which five systematic steps were developed to validate the findings of this study. Firstly, the search queries explored among the Thomson Reuters Web-of-Science and Elsevier Scopus databases, in the first step. In addition, the so-called grey literature - produced on government, business, academia and consulting firms in both electronic version and print format but not controlled by commercial publishing – was systematically reviewed (e.g. documents of international organizations, research reports etc.) In the second step, the duplicated articles found in both databases were eliminated. In Step 3, the abstract and keywords of the found articles were precisely studied to reach the relevant literature and exclude the irrelevant ones. In Step 4, the current study database is formed that included the most relevant articles. In this step of the literature review, a special attention was paid on the role of employment (legal regulations) of the platform companies. (e.g. the EU does not intend to intervene in the employment regulation of platform workers employment regulation belongs into the authority of the national legal systems of the EU member states.). The employment regulations dimension of platform working often missing in both the academic and grey literatures. In the final step, the articles precisely read, and finding, reports, and tables are presented accordingly.

**iii. Platform-based work – theoretical bases and terminological debates**

The platform-based economy is a relatively new phenomenon, there is an abundance rather than a deficit of definitions of platform-based jobs and companies in the social sciences. There have been several attempts to define platform-based work in recent years. For example, Eurofound uses the following concept of platform-based work: "Platform work is a form of employment that uses an online platform to enable organisations or individuals to access other organisations or individuals to solve problems or to provide services in exchange for payment." (Eurofound, 2018: 9)

The Eurofound study draws attention to the following characteristics to demonstrate the dynamic existence of platform-based work:

- Paid work is organised through an online platform.
- Three parties are involved: the online platform, the client, and the worker
- The aim is to carry out specific tasks or solve specific problems.

- The work is outsourced or contracted out.
- Jobs are broken down into tasks.
- Services are provided on demand (Eurofound, 2018: 9).

The conceptual and emerging cultural context is important to underscore: platforms are a result of digitization. The growth of digital technology in economic activities is meant by digitization. Digitization affects all markets, but not to the same degree. The digital economy is the portion of the economy that is solely reliant on digital computing technologies. Digital innovations not only alter the way we work, but they also contribute to a whole new business model in several cases. The platform economy, in which companies undertake to link supply and demand by building digital platforms in a product or service sector, is one manifestation of this.

We should also draw attention to the usage of the word, and the sector-specific characteristics of platform companies: "… Thus, service platforms such as Uber or Airbnb connect seekers with providers of services, goods platforms like Amazon connect buyers with sellers, and information platforms like Facebook and Google connect people to each other, to advertisers, and to information" (Rahman and Thelen, 2019: 179).

We will only deal with the platforms that connect the demand and supply side of the global labor market below. While outsourcing their operations that do not involve key competencies is not a new development for businesses, it is in all respects that they do so in an online global labor market. Amazon was one of the pioneers in the process, providing a template for coordinating purchasing, while streamlining related activities, such as translation, the wording of smaller documents, and customer service. Encouraged by the model's popularity, the same service was made available to other businesses, making Amazon Mechanical Turk one of the largest online labor

Electronic copy available at: https://ssrn.com/abstract=3764851

markets in the world.

Like the labor market overall, the online market is also segmented. These job placement platforms specialize in various activities, tasks with distinct skills that have serious repercussions not just for salaries, but also for the structure and functioning of the platform itself.

It is important to remember that the platforms are run by so-called algorithm-based management, the key characteristics of which are some or all of the following features:

- Prolific data collection and surveillance of workers through technology;
- Real-time responsiveness to data that informs management decisions;
- Automated or semi-automated decision-making;
- Transfer of performance evaluations to rating systems or other metrics; and
- Use of "nudges" and penalties to indirectly incentivize worker behaviors (Mateescu and Nguyen, 2019: 3).

**iv. The multidimensional nature of platform-based work**

One of the most significant aspects of platform-based work is its diversity, in which the structure of activities, qualification criteria, and the nature of the service is the most relevant segmenting variables. In relation to the above, Codagnone et al. (2016) and Pajarinen et al. (2018) differentiate platforms that offer personal/physical services (e.g. Uber, Bolt, AirBnB, Wolt, Delivero, etc.) from non-personal digital networks (e.g. Upwork, Guru, Cloud Factory, Amazon Mechanical Turk, etc.). There are two kinds of labor markets in this context: online labor markets, where the service is delivered in digital form, and mobile labor markets, where a personal presence is needed for the service (Pajarinen et al., 2018: 5). Depending on the length of employment and the need for skills, services in both markets can be further distinguished. For example, because of their routine nature, micro-tasks carried out on the online market (such as invoice processing) are relatively simple and fast to complete, while project-type, shorter-term jobs require higher-level skills and creativity (such as creating artificial intelligence-based algorithms). Similarly, there are several categories of tasks typical of the mobile market: on the one hand, personal services with minimum or medium qualifications are available (such as window cleaning), but there are also interactive services that require high qualifications, such as language teaching (Codagnone et al., 2016: 7 In: Pajarinen et al., 2018: 5). This labor market typology is presented in Table 2.

**Table 2. Types of labor markets and platform works**

|  | Online Labor Market (OLM) | | Mobile Labor Market (MLM) | |
|---|---|---|---|---|
| Service characteristics | Electronically transmittable tasks | | Services requiring personal presence | |
|  | Microtasks | (Mini) Projects | Physical services | Interactive services |
| Duration | Short | Long | Short | Long |
| Skill level | Low-to-Middle | Middle-to-High | Low | High |
| Dominant form of transactions | Peer-to-Business | Peer-to-Business | Peer-to-Peer | Peer-to-Peer |
| Examples | Amazon Mechanical Turk | Upwork | Uber | TakeLessons |

*Source:* Codagnone et al. (2016:7) and Pajarinen et al. (2018:5)

Pongratz (2018) adds two additional analytical dimensions to this classification that are relevant from the perspective of the heterogeneity of platform work. The first dimension is the average payment level and the other is the term used to describe workers, jobs, the platform itself, and the clients. Analyzing the content of 44 website operating platforms, Pongratz (2018) found that different types of platform work involve clearly different "discursive constructions" and that these discursive constructions exercise a strong influence as to how the partners involved (clients, workers, and the platforms) perceive themselves, the other partners and the work itself. Table 3 presents platforms that vary greatly from each other, primarily in terms of the characteristics of the job (such as difficulty, length, etc.) and the employment status of the platform staff, which is expressed in the language used in the platform.

**Table 3. The main types and semantics of various platforms**

|  | **Microtask** | **Freelance platforms** | **Specialized platforms** |
|---|---|---|---|
| **Task complexity** | Low | High | High |
| **Payment** | Low-paid | Higher wages | Higher |
| **Workers are addressed** | ... as workers | ... as freelancers | ... as freelancers |
| **Jobs are labelled** | Task | Project | Varies according to the purpose (design, translation, etc.) |
| **Platform designation** | Platform or marketplace | Platform or marketplace | Platform or marketplace |
| **Buyers are called** | Customers, clients, buyers | Customers, clients, buyers | Customers, clients, buyers |

*Source:* based on Pongratz (2018:63-64).

Jobs that are relatively easy and require low skills, for example, are usually referred to as" microtasks", while jobs that require medium or high skills are often characterized as" projects". Those with so-called microtasks that require low skills and provide minimum wages are usually referred to as workers, whereas freelancers or entrepreneurs are considered to be those who perform high-skilled jobs via a platform. Platform-based work is maximally individual in nature, and its employment characteristics only guarantee minimal employment security and protection in exceptional cases.

**v. Regulatory problems associated with platform-based work-Uber's failed launch in Hungary**

The well-known mantras of Silicon Valley are a clear indicator of the strategy pursued by platform companies in their business acquisition/establishment philosophies: "Don't ask permission, ask forgiveness!" or "Walk quick-walk the unknown route!" (Thelen (2018); Grabher and Tuijl, 2020). In bilateral markets, platform companies tend toward monopolistic "winner takes all" competitive practices. The first entrants to the market are quickly gaining leadership[6], enticing large masses of users from other platforms. The best-known example of this type of market dynamism is Facebook: after becoming an unquestionable market leader, consumers abandoned similar platforms. The well-known credo of market acquisition can be seen in the rapid, often aggressive action of platform companies: they consciously plan to overwrite the rules, but most often they try to take advantage of the loopholes in the existing rules. In certain cases, the current regulations on their operations are immature or non-existent, and it is difficult for national and foreign regulatory agencies to keep up with the speed of technological change and 'innovative' business activities (such as tax payments) that bypass regulation.

In the relationship between institutions and market participants, the interrelationships between hard and soft regulators[7] are well illustrated by the unsuccessful launch of Uber in Hungary.[8] The business was founded in San Francisco in March 2009,

---

[6] Although there are some exception. For example, Facebook wasn't the first entrant—My Space had been around for some time before. Likewise, for Google, Netscape and AltaVista came before but, in the end, couldn't compete. And only now, 20 years after Google won the browser wars outside of China, is the EU and US starting to address the monopoly of Google.

[7] For more details on the content of 'hard' and 'soft' regulators see Makó et al. (2020).

[8] The source of the empirical results collected by the qualitative research method used in the study is the Industrial Relations and Social Dialogue in the Age of Collaborative Economy (IRSDACE) international project, 2017–2018. The central aim of the research is the role of traditional advocacy institutions (e.g. trade unions, employers' associations) in the rapidly developing digital labor market, especially social dialogue in seven European countries (Belgium, Denmark, France, Hungary, Germany, Slovakia, Spain). The Hungarian case study prepared within the framework of the project examined the operation of three platforms

operating in the passenger transport sector. Within a few years, Uber had expanded to major cities across the globe, including Budapest, where it appeared as a market force in November 2014. The business model represented by Uber sparked a vigorous discussion on two main topics among social and economic actors in the Hungarian capital: the fact that the company did not pay most of the normal taxi tax for its passenger transport operations in Hungary,, and concern that Uber had an unfair competitive advantage over other companies that comply with the rules regulating passenger transport because of the platform-based business model. Uber claimed that was not a taxi company and employer but simply, a platform operator, a firm that develops high-tech apps and websites that connects travelers who require a service with drivers who provide the service. Rival taxi companies that "played by the rules" protested, citing the facts that:

1: Uber did not pay a mandatory deposit to the public passenger transport regulator like other taxi companies do.

2: Uber did not adhere to strict environmental regulations, claiming that it does not have a car fleet of its own.

3: Uber workers did not have to take an advanced level examination in traffic and driving, did not have to take a career aptitude test, and were exempted from taking a taxi business course, unlike their counterparts in a conventional passenger transport company.

4: Additionally, cars driven by Uber drivers were much easier to maintain than for other drivers because there was no requirement for an annual roadworthiness test, nor did they have to be covered and pay higher insurance rates.

---

(microwork, Airbnb and Uber) (Kahancová et al., 2020; Meszmann, 2018). Moreover, our own experience was also used regarding the operation of the business.

Uber's unwillingness to identify itself as a taxi company and to assume the responsibilities and financial pressures associated with it was the key source of dissatisfaction and conflict. Similarly, to the operating features of most platform companies, Uber, as a passenger transport platform regarded its service as" the neutral intermediary that solely matches the supply of and demand for independent contractors. By emphatically maintaining this claim (through a multitude of litigation cases across a multitude of jurisdictions), platform operators seek to avoid basic entitlements resulting from employment contracts such as social security, minimum wages as well as work time and security regulations" (Grabher and Tuijl, 2020: 1012). In Hungary, this picture is worth nuancing since in fact most conventional taxi companies often do not employ drivers, who are usually considered entrepreneurs. It should also be recalled that in Hungary, most foreign corporations do not pay part of their income, so Uber was not the only one.

Taxi drivers represent a traditionally well-organized interest group in Hungarian society and found a significant supporting partner in the Hungarian state in the conflict around Uber, mainly due to the issue of tax evasion. Meanwhile, Uber executives argued that they use a more transparent payment system than anyone else in the passenger transport sector, which is typically considered as a gray zone. This was also true as Uber operated its own system based entirely on electronic payment. A demonstration was organized by the taxi union, Uber was petitioned, and a similar campaign was run by the taxi companies' owners. A protest in the capital was organized by the Trade Union of Hungarian Taxi Drivers in January 2016, causing major traffic difficulties. Following the protest, a new law on passenger transport was passed by Parliament, which prohibited the type of service that Uber was intending to implement in the Hungarian capital. On 13 July 2016, the US taxi company left the country.

14Electronic copy available at: https://ssrn.com/abstract=3764851

Uber began to operate in Hungary like elsewhere in the world: it "pushed the door" to the market, but its innovative business model was totally in conflict with the regulation of the industry. It was soon possible for taxi drivers with good lobbying skills to make the government an ally, and the loop started to crowd around Uber. The company attempted to conform to the rules at that stage. It began demanding its drivers to take the appropriate training, according to Zoltán Fekete, Uber's head of operations in Hungary, and by February 2016 all their drivers became accountable (Magyar, 2016), but still maintained its opinion that the old regulations do not apply to them or would incur unnecessary expenses in their case. They initiated a dialogue on these issues, but traditional taxi companies successfully defended their interests, and the government refused to enter into negotiations, saying that in an otherwise rather regulated market, all companies must compete on equal terms regardless of the technology used.

The legislature introduced a new taxi law in the summer of 2016, which, however, made it entirely difficult for Uber to operate, so the company left Budapest, where it has not returned to date.

It should be noted that Taxify (formerly called Bolt), a platform-based service using digital platform and apps similar to Uber, has been running smoothly since the withdrawal of Uber, having acknowledged the regulatory conditions imposed by the state, and has been one of the most dynamically expanding taxi companies in Budapest until the beginning of the pandemic.

In the context of a wider international institutional-regulatory comparison, Thelen (2018) analyzed Uber's market entry in the United States, Germany, and Sweden and found substantial variations between the attitudes of service users (passengers), rivals, as well as social and economic actors that affect their relationships (such as government, employers' organizations, etc.).

He argues that Uber has managed to recognize itself in the U.S. "as a supporter of the free market and customers" (Thelen, 2018: 999), thereby establishing an alliance with both customers and the political elite[9]. Taxi companies in Germany, on the other hand, rapidly formed a coalition of interests with public transport companies, labor unions and other social actors. As a result, they have been able to take collective measures against Uber to "protect consumer interests" before it could gain serious loyalty among consumers with the lower prices and more flexible services of the taxi company. The Swedish situation also differed from the American and German cases: it reflects a regulatory trend in which "taxation has become the focal point of central legislation and has galvanized the collective action taken by traditional taxi firms, unions and the state. "The reason for the joint action was to comply with the norm of the Swedish social security system in the spirit of fairness in the form of tax payments" (Thelen, 2018: 949).

In Slovakia, yet another scenario played out., Uber also faced opposition from rival taxi companies there, too, but due to other localized factors, the case was subjected to simplified court proceedings in which a binding decision could be issued by the court in the out-of-court proceedings. In March 2018, the court ruled that Uber's activities breached existing laws, the company had no excuse not to comply with the general taxi company rules, and the company thus withdrew from Slovakia. The taxi drivers then turned their attention to Taxify/Bolt, and in court, they had a similar argument. They were not so fortunate this time, though, that the court did not find the comparison between the two taxi companies well-founded. As a result, a paradoxical situation arose: Taxify/Bolt was able to continue to run smoothly while Uber's operations were rendered impossible. The solution to the situation was waiting for the legislature: in November 2018, a new law filled the legal gap. In the spring of 2019, Uber returned to Bratislava undisturbed (Martinek, 2020).

---

[9] However, there has been a major change of focus on the safety of the rights of workers in the institutional climate of the United States. The California parliament, for instance, required state-based platform-based taxi companies (Uber, Lyft, and Doos Dash) to pay sick leave and rest periods to their employees in September 2019 (Conger and Scheiber, 2019).
16
16Electronic copy available at: https://ssrn.com/abstract=376485116

The case of Uber in these various countries exemplifies the central role and complexities played by institutions in the social acceptance and effect of technical and business model innovations.

**vi. Dilemmas of labor law and civil law regulation[10]**

Legal regulation is in theory the most important institutional factor that can set the working conditions of platform workers. Platform workers are considered primarily independent contractors under Hungarian labor law, and self-employed person work is regulated by civil law However, civil law, as opposed to labor law, does not offer job protection. Hungarian labor law is relatively unprepared for dealing with platform-based jobs. Recent studies of the practice of platform work have shown that" Hungary is characterized by the early development stage of the so-called 'gig economy,' which is immature both in terminology and regulation. In addition, platform work is a scarcely visible and marginal type of work, and social partners are less aware of it and do not treat it as an employment phenomenon requiring special

---

[10] The chapter of labor law regulation is partly based on the contribution of labor lawyers Tamás Gyulavári (Pázmány Péter Catholic University, Department of Labor Law) and Zoltán Bankó (University of Pécs, Faculty of Law). We would thereby like to thank the authors for their work. A more detailed exploration of the problems of platform employment from a labor law perspective can be found in Rácz, I. (2020).

regulation. Working on platforms is not a topic of discussion. There are no regulatory or legal development efforts in Hungary aimed at protecting the working conditions and/or social protection of platform workers" (Kun and Rácz, 2019: 10)

Experience from the recent platform-based work survey of trade unions indicates that most trade union leaders consider the future role of trade unions to be significant. In addition, supporting new social movements and alliances is considered important by some trade union confederations, such as the Intellectual Trade Union Confederation (Borbély et al., 2020]). In Hungary, a very rigid, so-called binary model rule the legal approach to employment relations, reflecting the duality of labor law and civil law control, and which implies either "universal" or "zero" legal protection.

In the binary control approach, the platform worker enjoys full employment protection in possession of 'employee status' - after being protected by the rules of the Labor Code - or in 'self-employed' (entrepreneurial) status without any legal rights and is regulated by the Civil Code. Currently, there is no special legislative regulation for the third group of workers in the Labor Code. However, the concept of an economically dependent worker or subcontractor, which can not conform to the standard and atypical types of employment recognized under Hungarian labor law, is also recognized in some countries: in Italy (parasubordinati), for example, or in Spain (TRADE), in the UK (worker) but a similar semi-employment status exists in Germany as well.

The so-called algorithm-based management, which is most achievable in the framework of assessment and pricing, is one of the unique features of platform-based work. The service provider is assessed by the service customer during the evaluation process. For instance, the five-star rating system, now commonplace, promoted by the Amazon website, has become widespread in many customer ratings of service, such as Uber (Chan, 2019) and its competitors, which has led to concerns of unfair pressure on workers, who, as we examine below, stand to lose in the "ratings game" with no recourse to bad customer reviews that drop their ratings below near-perfect scores. The Job Success Score (JSS) is a more sophisticated method used by Upwork: the client assesses the successful project-and therefore the service provider. Evaluation is often a source of confidence in the service provider by the customer/client, but the lack of transparency in assessment inevitably leads to problems. For example, if a service provider does not agree with a service user's rating (score), they would be unable to negotiate and find a consensus in most situations because there are no channels where they can make their voices heard. However, below a certain score or based on unfavorable feedback from service customers, the platform automatically disconnects from the service provider. All this puts platform workers struggling to satisfy the dictates of the algorithm in a vulnerable position, especially in terms of bargaining power. The timeliness and relevance of this subject are well illustrated by the recent re-emphasis of the Nordic countries leading digital work on key features of platform work, such as transparency, fairness, accountability, benefit-sharing and learning, and innovation (Seppanen and Poutanen, 2020).

The problems of algorithm-based management and digital assessment are almost entirely absent from the legislation of Hungarian labor law, rendering it difficult to call out the flaws and potential correction of online service evaluation. The transfer of assessments from one platform to another is still an unresolved concern in the absence of legal guarantees. Disciplinary sanctions or the severance of legal relationships are the two potential effects of online evaluation. For example, in the case of a taxi platform operated by Taxify/Bolt, the passenger evaluates the driver's work on a scale of 1 to 5. The latter does not know the evaluation aspects of the passenger, the passenger evaluates anonymously, therefore it is extremely difficult to modify the unfavorable evaluation and have a subsequent correction. An even bigger problem is that the platform operator asks the passenger to evaluate the overall quality of the trip ("How was your trip?"), not just about the job of the driver. Consequently, if there are issues, for example, the mobile application underestimates the waiting time before the taxi arrives or incorrectly locates the geographical position of the passenger (these are daily issues)-the disappointment of the passenger, the low evaluation is expressed in the evaluation of the driver and not in the business[11].

19Electronic copy available at: https://ssrn.com/abstract=3764851

Drivers are underrated in these situations because of conditions on which they have little control. However, evaluations can have severe consequences. For example, in the case of the Bolt, if the driver's assessment after 40-50 occasions is lower than a certain score, the relationship to the platform is immediately lost, which is a big psychological burden for the underrated, often innocent drivers. But the passenger is also unable to directly contact the company (e.g. the department of customer relations) and only has the ability to send feedback through a different digital platform share (Appstore) should they have a company (and not the driver) -related problem. The textual evaluation already mentioned is also an alternative, but it is a one-way communication that does not provide the passenger with any details about what steps have been taken to deal with the issue.

According to the Labor Code, as provided for in a collective agreement or employment agreement, disciplinary sanctions may be applied. The Civil Code, on the other hand, allows the parties involved to agree on the consequences of a disciplinary violation. In contrast with a "simple" termination of employment,

---

[11] Of course, it is also possible to evaluate the journey in text, but this is much more difficult than giving stars.

platform workers are generally deprived of all protection.

Another issue beyond the scope of the Labor Code is contractual rights, particularly the right to conclude a collective agreement. The scope of the collective agreement in the Hungarian labor market is almost entirely focused at the level of the workplace. If at least 10 percent of the employer's workers are labor union members, a collective agreement can be concluded by a trade union or trade union organization. Since the vast majority of platform workers do not have official employee status, collective agreements do not apply to them. For the atypical types of employment that differ from conventional employment forms, such as platform-based work, an industrial collective agreement may also be an ideal solution. In Hungary, however, industrial collective agreements are unusual. Middle-level social dialogue and the work of the dialogue committees are governed by Act No. 74 of 2009 on Industrial Social Dialogue[12], but the Act only applies to the representation of the interests of those with standard employee status. European Union competition law rules banning the conclusion of a collective agreement with workers who do not have a traditional status as an employee also appear in the Hungarian legal regulatory procedure.

Hungarian labor law does not, however, represent issues relating to the protection of platform workers. Creating separate and detailed legislation for the new type of work/employment is an urgent need, according to labor lawyers. Comparing the regulatory differences between so-called offline and online employment would provide a better understanding of the content of radical changes affecting platform economy employment. In this context, it is worth highlighting the evaluation of Grabher and Tuijl (2020) according to which "…platforms accelerate the 'vanishing of the corporation' and expedite the secular shift from (long-term) employment relations over (short-term) jobs to (discrete) gigs. This 'taskification' of work transforms professional careers governed by (offline) accumulated human capital into contractual portfolios shaped by (online) reputation capital" (Grabher and Tuijl, 2020: 1012).

---

[12] Act LXXIV/2009 on Sectoral Dialogue Committees and Certain Issues in Mid-Level Social Dialogue (https://net.jogtar.hu/jogszabaly?docid=a0900074.tv).

In connection with the employment conditions of platform workers and the representation of their interests, we should refer to the so-called new global initiatives on 'soft regulation' in addition to the tough labor rules. The "Platform for a New Economy and Society" of the World Economic Forum accepted the Charter of principles for good platform work in 2020. The eight principles include promoting the participation and advocacy needs of platform workers, as well as equal working conditions, social security, decent earnings, learning and development opportunities as well as data management (WEF, 2020).

**vii. Conclusions and future research challenges – the" societal impact"**

The semiotic clarification of concepts related to platform-based work is not a needless theoretical activity: the terminology used in the research has a huge impact on how we examine this phenomenon and the social and economic regulation around it (Kenney and Zysman, 2016). We used a multidimensional concept of platform work in our paper, which includes the complexity of tasks (micro-versus project work), the amount of income (high versus low), and types of employment status (contract worker versus freelancer/entrepreneur). The diversity of platform work is well illustrated by the Upwork global platform, which has more than a dozen (13) professional profiles. In comparison to the minimum skills needed for micro-tasks, disparities in platforms (such as the innovation and skills required for micro-project work) produce significantly different negotiation positions and complex advocacy needs for platform staff.

In the absence of uniformly used terminology, as well as due to the rapidly changing dynamics of this form of employment, estimating the proportion of platform workers working in the European Union is particularly challenging. Another difficulty is the unequal distribution of research in different groups of countries in the European Union. We have a relatively rich research experience in the continental, Anglo-Saxon and Nordic countries, as opposed to the Mediterranean group or the new member states. There have been several major surveys in the EU over the last half-decade, for example about 2017–2018: COLLEEM (Pesole et al., 2018, 2019), 2016–2019: (Huws et al., 2019), 2018–2019: ETUI (Piasna and Drahokoupil, 2019).

These surveys provide important, albeit incomplete, snapshots of key features of platform work. Despite the difficulties indicated, we have reliable data that currently a relatively small part of the European workforce - about one-tenth - participates in the digital labor market, with varying degrees of intensity.

The most common features of platform-based work are discussed in these detailed EU surveys but do not include details on its complex nature and its integration into socio-institutional settings. The diversity of the task structure, issues of transparency and justice, employment status, and advocacy in the conflict resolution between service providers and consumers are rarely accurately examined. Some of the few exceptions include, for instance, Thelen (2018), Grabher and Tuijl (2020), as well as Greber and Krzywdzinski (2019).

In addition to important statistical analyses of European surveys, we have applied a case study-based methodology to identify the complex factors of radical changes produced by platform companies. The emergence of the world-leading Uber platform-based business model in the passenger transport sector has drawn the interest of both theoretical and practical professionals. Both successes and setbacks followed its appearance. For instance, the company was forced to face the collective activity of the taxi society in large German cities, similar to the Hungarian capital. Mainly because of the unified action of taxi drivers and the Hungarian legislature -taking the revenue before tax of Uber abroad- the decision was taken in favor of taxi drivers. In July 2016, Uber withdrew from Budapest after the company opted not to follow the Hungarian regulations. Bolt/Taxify, a company with a similar profile focused on the digital platform business model, filled the sudden vacuum in the capital's passenger transportation market following Uber's departure.

In both the offline and online labor markets, Hungarian labor unions have exceptionally weak negotiating positions. In general, there is a modest interest in the recruitment of so-called precarious workers.[13] The future regulation of platform-based employment needs to play a vital role in improving the bargaining position of Hungarian labor relations partners, which traditionally have been weak. Resolving the dilemmas of labor law and competition law requires further social dialogue between actors in different positions in industrial relations, based on empirically collected and evaluated data. In addition to the traditional actors in industrial relations, such as employers 'and trade unions' associations, and relevant government labor entities), we need to draw attention to the role of emerging new institutions addressing the concerns of platform workers (such as Sharing Economy Association – https://www.sharingeconomy.hu).

---

[13] For example, according to a survey conducted relatively long ago - but to date on the only comprehensive, representative national sample (2010)-, the vast majority of Hungarian trade union leaders (78-89 per cent) are not interested in precarious employment in the traditional (offline) labor market: workers (such as part-time, fixed term or temporary workers). (Neumann, 2018: 81) The role and attitudes of trade union leaders towards platform workers in the digital labor market are even less identifiable.

Their role and operation may be one of the most exciting research issues in the field of labor relations in the coming years. Based on the prevailing theoretical approaches to the platform economy and European empirical research, studies that explore the content, employment status, and advocacy issues of platform-based work from a comparative perspective of the offline and online labor markets may be of primary interest to theoretical and practical experts. In this context, we wish to emphasize the research experience according to which platforms are not neutral intermediaries, but organizations that play an active role in shaping the digital workflow framework and working conditions. They encode and display in the form of visibly neutral technical frameworks the power relations that underpin the relationship between capital and labor (Greber and Krzywdzinski, 2019).